\documentclass[conference,a4paper]{APSIPA2021}
\usepackage{multirow}
\usepackage{amsmath}
\usepackage[psamsfonts]{amssymb}
\usepackage{amsxtra}
\usepackage{threeparttable}

\usepackage{kotex}
\usepackage{graphicx}
\usepackage{multirow}
\usepackage{tabularx}
\usepackage{booktabs}
\usepackage[export]{adjustbox}
\usepackage[hyphens,spaces,obeyspaces]{url}

\usepackage{geometry}
\begin{document}

\title{CochlScene: Acquisition of acoustic scene data using crowdsourcing}

\author{%
\authorblockN{%
Il-Young Jeong and
Jeongsoo Park
}
\authorblockA{%
Cochl, Inc., Seoul, Republic of Korea \\
E-mail: \{iyjeong, jspark\}@cochl.ai
}
}

\maketitle

\begin{abstract}
This paper describes a pipeline for collecting acoustic scene data by using crowdsourcing. The detailed process of crowdsourcing is explained, including planning, validation criteria, and actual user interfaces. As a result of data collection, we present CochlScene, a novel dataset for acoustic scene classification. Our dataset consists of 76k samples collected from 831 participants in 13 acoustic scenes. We also propose a manual data split of training, validation, and test sets to increase the reliability of the evaluation results. Finally, we provide a baseline system for future research.
\end{abstract}

\section{Introduction}

Acoustic Scene Classification (ASC) is the task of analyzing audio signals recorded from a specific location to classify its acoustic environmental context. It is an important task in itself, but it is also helpful to utilize contextual information in various acoustic analysis tasks such as speech recognition or security surveillance. Therefore, ASC has attracted many researchers in the field of audio signal processing and machine learning. \cite{Abeber2020, Barchiesi2015, Han2017}.

Despite the importance of ASC, the difficulty of collecting data has been one of the bottlenecks of the studies. Unlike acoustic event data (\textit{e.g.} gunshot or glass break) which can be generated from multiple objects in the same place, acoustic scene data needs to be collected from as many different places as possible to maximize its diversity. In addition, since this acoustic context would be highly related to culture and country-specific factors, data collection across a variety of geographically distinct sites would be desirable.

One of the most well-known datasets for ASC is the one used for the Detection and Classification of Acoustic Scenes and Events (DCASE) challenge~\cite{Dcaseurl}, an annual event for various tasks in environmental audio analysis, including ASC. The ASC-related task and dataset in DCASE have been changing each year, considering the usability in real-world situations, including using different recording devices, external data, unseen classes, low-complexity models, and visual cues~\cite{Mesaros2016, Mesaros2017, Mesaros2018, Heittola2020}. It has succeeded in attracting the attention of those looking for the ASC dataset as well as those who want to participate in the challenge. Besides DCASE, several datasets, including LITIS~\cite{Rakotomamonjy2015}, have been used.

Instead of collecting acoustic scene data manually by themselves, one can try to collect data from volunteers or by using crowdsourcing.\footnote{Crowdsourcing can be applied not only to the data collection process but also to the labeling of collected data, including tagging, segmentation, transcription, etc~\cite{Eskenazi2013, Buhrmester2016, Lee2022, Park2021}. In this paper, we focused on the data collection process only.} For example, Freesound is a collaborative platform for collecting various audio samples. Several datasets for audio event detection and classification have been released as a subset of the Freesound database, including FSD50K \cite{Fonseca2021}. Although it has successfully collected an enormous amount of audio data efficiently, its open platform made it difficult to collect a sufficient quantity of high-quality targeted data efficiently.

This paper introduces Cochl Acoustic Scene Dataset, or CochlScene, a new acoustic scene dataset whose recordings are fully collected from crowdsourcing participants\footnote{\url{https://zenodo.org/record/7080122}}. Most of the initial plans and guidelines for the processes were provided by the researchers in the field of audio signal processing and machine learning, including the authors. Selectstar\footnote{https://selectstar.ai}, a data crowdsourcing company in Korea, performed the actual process and modified the initial plan through discussion, considering the difficulty in the actual process. As a result of extracting a subset related to ASC from the entire collection, we collected 76,115 10 seconds files in 13 different acoustic scenes from 831 participants. 

The rest of the paper is organized as follows. In Section II, the collection procedure is described in detail, and CochlScene dataset is introduced as a result in Section III. In Section IV, we provide the baseline system for ASC using CochlScene and discuss its results. In Section V, we conclude the paper and discuss future work.

\section{Crowdsourcing for Acoustic Scene Data Collection}

\subsection{Target Class Setting}

We first selected the target acoustic scene classes and added/removed some during the collection process. We mainly considered the following:

\begin{itemize}
\item \textit{\textbf{Participants population}} We focused on the acoustic scenes that many participants could conveniently visit. Therefore, we chose the target classes in urban areas and excluded some nature scenes (\textit{e.g.} Moutain, Sea, Waterfall).
\item \textit{\textbf{Cultural bias}}  Since a company in Korea performed the crowdsourcing process, we expected that the participants were mostly located on the same specific cultural basis. We tried to avoid this bias in the dataset and excluded scenes which are uncommon in Korea (\textit{e.g.} Tram) or rare outside of Korea (\textit{e.g.} Traditional Korean Palace). 
\item \textit{\textbf{Ambiguity}} During the planning and actual collection process, we found out that some acoustic scenes are difficult to be defined (\textit{e.g.} Parking lot) or hard to be expected to have meaningful acoustic cues (\textit{e.g.} Bakery or Florist). We excluded such scenes to eliminate ambiguity.
\item \textit{\textbf{Accessibility}} The places that require permission to enter and record sounds (\textit{e.g.} Factory, Spa) are not suitable for the crowdsourcing-based collection since it is not available for unspecified participants. Inducing the participation of people who can access this place by increasing the reward is another option, while we just excluded those classes in this work.
\item \textit{\textbf{Submission rate}} Although it passed all the above considerations, some classes (\textit{e.g.} Bus station) had a relatively low submission rate than others. We stopped collecting those classes if it seemed they could not reach the target quantity.
\item \textit{\textbf{Acoustic similarity}} Some scenes might be semantically similar (\textit{e.g.} Cafe and Restaurant), or have similar acoustic characteristics (\textit{e.g.} Street and Residential Area), while those were not excluded for this reason.
\end{itemize}

\subsection{Collection Platform}

\begin{figure}[t]
\begin{center}
\includegraphics[]{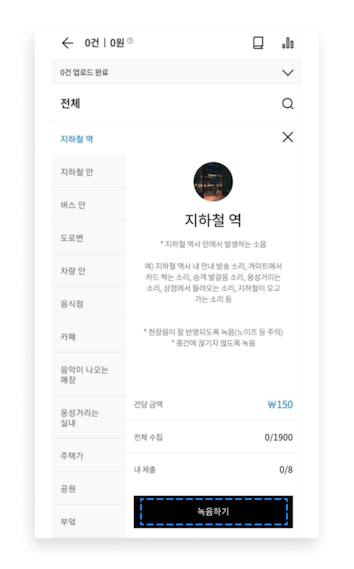}
\end{center}
\caption{A screen capture of the crowdsourcing application used for collecting acoustic scene data. The left menu lists the acoustic scenes, and the right part describes the description, reward, targeted amount, and the number of data the user submitted. Source:~https://selectstar.ai}
\label{fig:app_screenshot}
\end{figure}

A smartphone application developed by SelectStar was used for the participants to upload the recordings. Fig.~\ref{fig:app_screenshot} shows its basic interface. Most of the participants were existing users of the application who had experience participating in other collection tasks, including image and speech data. To submit the recordings, users could record the sound only through the application right before the submission, while uploading the existing files was not allowed to prevent manipulation. Each participant could upload a limited number of recordings for each class, which was slightly increased for the classes with a low submission rate.

\subsection{Data Validation}

We also used crowdsourcing to check the quality of the submitted data. 3 to 5 participants listened to each submission and made the binary decision on its quality, and the submission was accepted or rejected based on the majority voting. The following were considered, though these criteria would be subjective and limited as the validation was based solely on the audio recording and no additional information.

\begin{itemize}
\item \textit{\textbf{Recording gain}} The submissions whose gain was extremely low (not perceivable in normal volume) or high (causes clipping) were rejected.
\item \textit{\textbf{Privacy}} The submissions containing speech that could be easily identified or contained private topics were rejected. Some submissions, however, may contain speech from the participant who recorded under their agreement.
\item \textit{\textbf{Acoustic cue}} The submissions were rejected if no cue about its target scene was found from the recording. However, we did not restrict that these acoustic cues have to be only related to the target scene; for example, if the submitted data was estimated to be recorded either in Cafe or Restaurant when the target class was Cafe, it was accepted.
\end{itemize}

\subsection{Data Normalization}

All the accepted submissions were normalized to have the same format. First, We divided each recording into files of 10-second segments and discarded the remainders. It is noted that the acoustic cue in section II.C was considered for the entire recording in data validation, so each segment may not have a clear cue for the acoustic scene. Each file was saved with a file name of the form \{Class\}\_\{User ID\}\_\{Submission ID\}\_\{Segment index\} \textit{e.g.} Bus\_user0017\_14896147\_000.wav.

\section{Cochl Acoustic Scene Dataset}

\begin{table*}[]
\caption{\label{tab:dataset_stat}The number of files and participants in CochlScene.}
\begin{center}
\begin{tabular}{c|cccc|cccc}
\multirow{2}{*}{Class} & \multicolumn{4}{c|}{\textbf{Number of files (number of submissions)}}                                                            & \multicolumn{4}{c}{\textbf{Number of participants}}                                                                           \\ \cline{2-9} 
                       & \multicolumn{1}{c|}{\textbf{Total}} & \multicolumn{1}{c|}{\textbf{Train}} & \multicolumn{1}{c|}{\textbf{Val}} & \textbf{Test} & \multicolumn{1}{c|}{\textbf{Total}} & \multicolumn{1}{c|}{\textbf{Train}} & \multicolumn{1}{c|}{\textbf{Val}} & \textbf{Test} \\ \hline
Bus                    & 5,821 (950)                         & 4,653 (761)                         & 577 (92)                          & 591 (97)      & 172                                 & 122                                 & 19                                & 31            \\
Cafe                   & 5,867 (950)                         & 4,691 (762)                         & 583 (95)                          & 593 (93)      & 192                                 & 138                                 & 23                                & 31            \\
Car                    & 5,845 (950)                         & 4,676 (761)                         & 584 (95)                          & 585 (94)      & 180                                 & 129                                 & 17                                & 34            \\
CrowdedIndoor          & 5,835 (950)                         & 4,663 (756)                         & 580 (96)                          & 592 (98)      & 213                                 & 146                                 & 25                                & 42            \\
Elevator               & 5,874 (950)                         & 4,696 (760)                         & 582 (95)                          & 596 (95)      & 300                                 & 210                                 & 41                                & 49            \\
Kitchen                & 5,969 (950)                         & 4,774 (759)                         & 595 (96)                          & 600 (95)      & 270                                 & 195                                 & 27                                & 48            \\
Park                   & 5,744 (950)                         & 4,593 (759)                         & 570 (95)                          & 581 (96)      & 194                                 & 131                                 & 22                                & 41            \\
ResidentialArea        & 5,758 (949)                         & 4,603 (759)                         & 574 (95)                          & 581 (95)      & 233                                 & 168                                 & 31                                & 34            \\
Restaurant             & 5,933 (950)                         & 4,745 (760)                         & 590 (95)                          & 598 (95)      & 198                                 & 135                                 & 19                                & 44            \\
Restroom               & 5,930 (950)                         & 4,739 (760)                         & 591 (95)                          & 600 (95)      & 304                                 & 205                                 & 34                                & 65            \\
Street                 & 5,745 (950)                         & 4,594 (759)                         & 572 (95)                          & 579 (96)      & 216                                 & 154                                 & 25                                & 37            \\
Subway                 & 5,897 (950)                         & 4,714 (760)                         & 588 (95)                          & 595 (95)      & 166                                 & 118                                 & 19                                & 29            \\
SubwayStation          & 5,897 (950)                         & 4,714 (762)                         & 587 (93)                          & 596 (95)      & 191                                 & 137                                 & 23                                & 31            \\ \hline
Total                  & 76,115 (12,349)                     & 60,855 (9,878)                      & 7,573 (1,232)                     & 7,687 (1,239) & 831                                 & 423                                 & 118                               & 290          
\end{tabular}
\end{center}
\end{table*}

\subsection{Dataset}

CochlScene consists of 76,115 single-channel audio files among 13 acoustic scene classes, and each file has 10 seconds length in 44.1kHz sample rate. The detailed numbers regarding the CochlScene are listed in Table~\ref{tab:dataset_stat}. Each class has a similar but not identical number of files, from 5,744 to 5,969. These files were obtained from the original 12,349 submissions with varied lengths, from 50.68 to 301.45 seconds. 90.70\% of the submissions have a length between 60 and 70 seconds.

In total, 831 participants submitted their recordings. The number of submissions per participant varies, and 85 participants submitted one recording while the others submitted up to 103 each.


\subsection{Data partitioning}

To reduce the effect of skewed submission distributions on participants, we provide manual data partitioning for training, validation, and testing. The main aims are 1) to avoid the submissions from the same participants being included in more than one partition and 2) to maximize the number of participants in the test set to improve the reliability of the evaluation results.

 We first counted the number of files (10-second segments) for each user and sorted them. The user with the most files was assigned to training sets if any class did not exceed the target quantity after assigning. It continues for each user in order until no more users can be assigned to the training set. After that, the same process was repeated for the validation set using the unassigned data, and the rest was assigned to the test set. The target quantity for training and validation was 80\% and 10\% of the total files for each class, respectively.

\section{Baseline system}

\subsection{System overview}
The baseline model architecture is shown in Table ~\ref{tab:model_architecture}. The proposed baseline system is designed to have the same structure as DCASE 2022 task 1 baseline\footnote{https://github.com/marmoi/dcase2022\_task1\_baseline} \cite{Mesaros2018}. The differences can be found in the input and output shapes; we assume the input audio has a length of 10 seconds with 44.1kHz in sample rate, whereas the DCASE baseline model receives 1-second audio with the same sample rate. To recognize the 13 scenes in the dataset, we also modified the output shape of the Dense\_2 layer from 10 to 13. We also apply batch normalization, ReLU activation, and dropout like the DCASE model, even though they are omitted in the table. The model has a total of 116,821 trainable parameters.

To convert the audio input to a Mel-scale spectrogram, the Kapre library was used \cite{Choi2017}. The Window size and hop size were set to 40ms and 20ms, respectively, as in the DCASE model. In addition, 40-dimensional log Mel filterbanks were used with Kapre's basic Mel-scale conversion library. Note that the time and frequency dimensions are swapped using the Permute layer after the melspectrogram conversion. 

The Adam optimizer with a learning rate of 1e-3 was used for training.
The training procedure was stopped when the validation accuracy was not improved for ten successive epochs or when it reached the 100th epoch. The implementation of the baseline system is publicly available in Cochl's GitHub repository\footnote{\url{https://github.com/cochlearai/cochlscene}}.

\begin{table}[]
\caption{\label{tab:model_architecture}Baseline model architecture.}
\begin{center}
\begin{tabular}{lll}
\hline
\textbf{Layer} & \textbf{Output Shape}  \\
\hline
Input & (441000, 1) \\
Melspectrogram & (499, 128, 1) \\
Permute & (128, 499, 1) \\
Conv2D\_1 & (128, 499, 16) \\
Conv2D\_2 & (128, 499, 16) \\
MaxPolling2D\_1 & (25, 99, 16) \\
Conv2D\_3 & (25, 99, 32) \\
MaxPolling2D\_2 & (6, 4, 32) \\
Flatten & (768) \\
Dense\_1 & (100) \\
Dense\_2 & (13) \\
\hline
\end{tabular}
\end{center}
\end{table}

\subsection{Performance}
The overall performance of the baseline model, including precision, recall, and F-score, is summarized in Table~\ref{tab:testset_performance}. We found that the baseline model shows a worse F-score in Cafe, CrowdedIndoor, Park, and ResidentialArea classes while it performs comparably better for Car, Kitchen, and Restroom. We expect that the classes with higher F-score are easier to be classified because they may contain unique sound events, such as engine sounds for Car, chopping sound for Kitchen, and flushing sound for Restroom. For the classes with lower F-score, we assume several classes may have semantic or acoustic similarities, so each can be easily confused.

\begin{figure}[t]
\begin{center}
\includegraphics[width=\columnwidth]{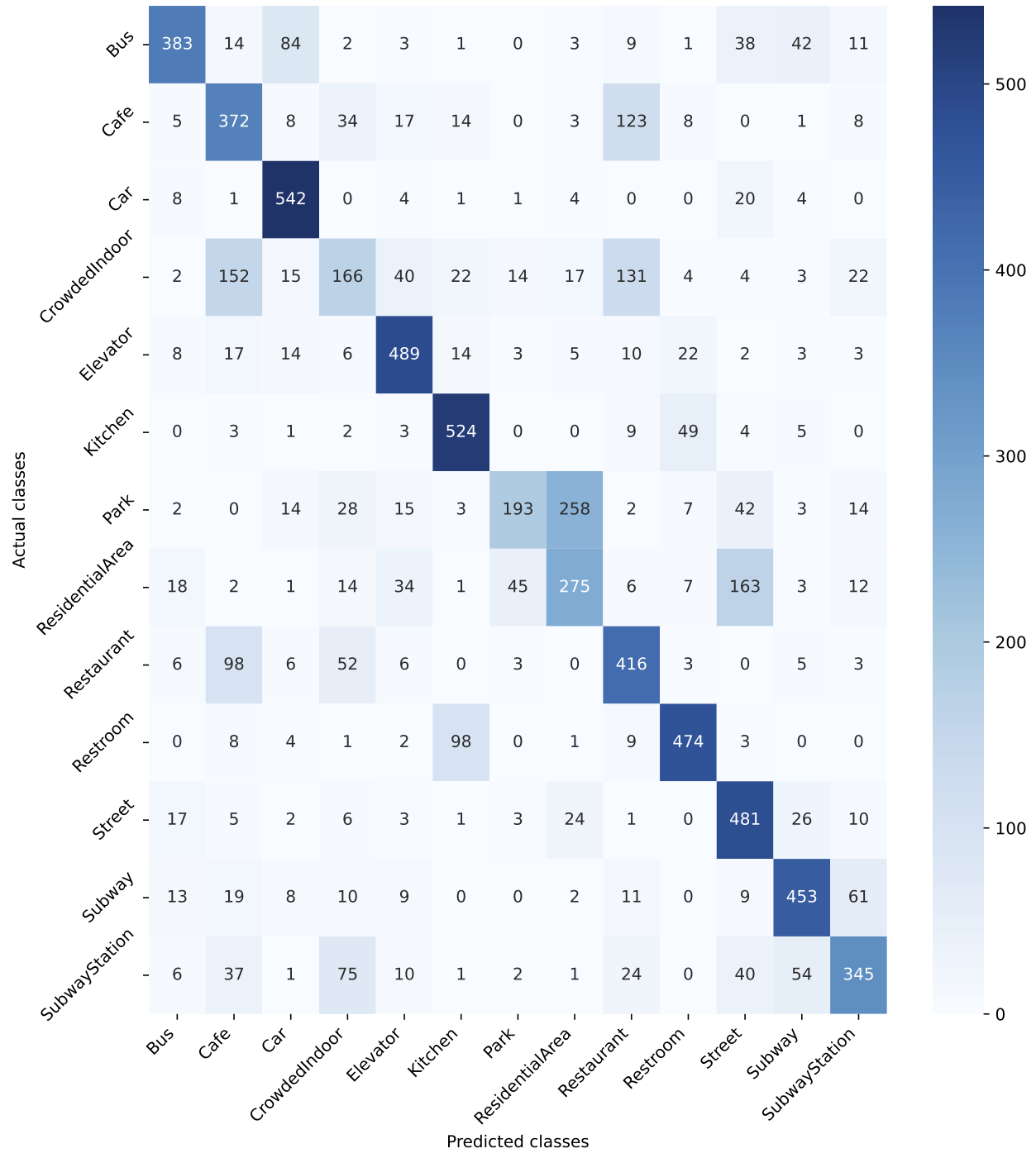}
\end{center}
\caption{Confusion matrix of the baseline model}
\label{fig:confusion_matrix}
\vspace*{-3pt}
\end{figure}

Fig.~\ref{fig:confusion_matrix} shows a confusion matrix of baseline model for a more in-depth examination. We found that Cafe, CrowdedIndoor, and Restaurant are likely to be wrongly recognized as each other. Besides, we can find asymmetries in the confusion matrix; for example, only 34 Cafe data are classified as CrowdedIndoor, whereas 152 CrowdedIndoor data are classified as Cafe. We can assume that Cafe data often contain certain clues that CrowdedIndoor data do not. For example, we can expect that sounds from a coffee grinder or espresso machine may rarely occur in the CrowdedIndoor data. Similar relations can be found between CrowdedIndoor and Restaurant because dish clanking and silverware rattling sounds could only be found in the Restaurant data. We can also find the asymmetric confusion among Park, ResidentialArea, and Street, where the confusion mostly occurs in the direction of Park$\rightarrow$ResidentialArea and ResidentialArea$\rightarrow$Street. We leave a detailed investigation of the acoustic characteristics of each class as future work.

\begin{table}[]
\caption{\label{tab:testset_performance}Test set performance of the baseline model.}
\begin{center}
\begin{tabular}{c|ccc}
\cline{1-4}
\textbf{Label} & \textbf{Precision} & \textbf{Recall} & \textbf{F-score} \\ 
\cline{1-4}
Bus	            & 0.818	        & 0.648	        & 0.723 \\
Cafe	        & 0.511	        & 0.627	        & 0.563 \\
Car	            & 0.774	        & 0.926	        & 0.844 \\
CrowdedIndoor	& 0.419	        & 0.280	        & 0.336 \\
Elevator	    & 0.770	        & 0.820	        & 0.794 \\
Kitchen	        & 0.771	        & 0.873	        & 0.819 \\
Park	        & 0.731	        & 0.332	        & 0.457 \\
ResidentialArea	& 0.464	        & 0.473	        & 0.468 \\
Restaurant	    & 0.554	        & 0.696	        & 0.617 \\
Restroom	    & 0.824	        & 0.790	        & 0.807 \\
Street	        & 0.597	        & 0.831	        & 0.695 \\
Subway	        & 0.752	        & 0.761	        & 0.757 \\
SubwayStation	& 0.706	        & 0.579	        & 0.636 \\
\cline{1-4}
Average          &0.669         & 0.664         & 0.655 \\
\cline{1-4}
\end{tabular}
\end{center}
\end{table}

\section{Conclusions}

Despite the importance of acoustic scene classification, collecting a highly diverse relevant dataset was a bottleneck of the studies. In this paper, we have provided an efficient approach to data collection based on crowdsourcing and released a new dataset as a result of this approach. A baseline system was also presented, and it is expected that other researchers can use it for their future studies.

We are planning future work to improve the efficiency and diversity of the dataset. First, we will investigate how to advance the role of crowdsourcing participants, including proposing new useful acoustic scene classes by themselves. In addition, considering that most of the participants are from the same country (Korea), one can expect a certain degree of culture-related bias from the dataset. For this reason, we plan to expand this pipeline for collecting data from participants in multiple countries to maximize the diversity of the dataset.

\section*{Acknowledgment}

We thank Selectstar for conducting the data collection process by using their platform as well as providing valuable comments.

\end{document}